\documentclass[Physsubmission, Phys]{SciPost}

\usepackage{blindtext,slashed,multicol,soul}

\hypersetup{
    colorlinks,
    linkcolor={red!50!black},
    citecolor={blue!50!black},
    urlcolor={blue!80!black}
}

\usepackage[bitstream-charter]{mathdesign}
\DeclareSymbolFont{usualmathcal}{OMS}{cmsy}{m}{n}
\DeclareSymbolFontAlphabet{\mathcal}{usualmathcal}

\newcommand{\calC}{\mathcal{C}}
\newcommand{\lagr}{\mathcal{L}}

\begin{document}

\begin{center}{\Large \textbf{
Global Fits of Dirac Dark Matter Effective Field Theories \\
}}\end{center}

\begin{center}
{\bf Ankit Beniwal\textsuperscript{$\dagger$}} (On behalf of the GAMBIT Collaboration)
\end{center}

\begin{center}
Theoretical Particle Physics and Cosmology Group, Department of Physics, \\King’s College London, Strand, London, WC2R 2LS, UK
\\[2mm]
\textsuperscript{$\dagger$}\href{mailto:ankit.beniwal@kcl.ac.uk}{ankit.beniwal@kcl.ac.uk}
\end{center}

\begin{center}
\today
\end{center}


\definecolor{palegray}{gray}{0.95}
\begin{center}
\colorbox{palegray}{
  \begin{tabular}{rr}
  \begin{minipage}{0.1\textwidth}
    \includegraphics[width=30mm]{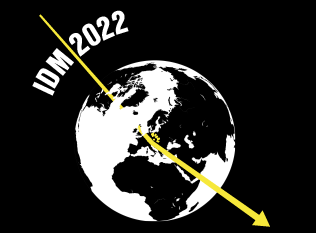}
  \end{minipage}
  &
  \begin{minipage}{0.85\textwidth}
    \begin{center}
    {\it 14th International Conference on Identification of Dark Matter}\\
    {\it Vienna, Austria, 18-22 July 2022} \\
    \doi{10.21468/SciPostPhysProc.?}\\
    \end{center}
  \end{minipage}
\end{tabular}
}
\end{center}

\section*{Abstract}
{\bf
In this proceeding, we present results from a global fit of Dirac fermion dark matter (DM) effective field theory (EFT) based on arXiv:\href{https://arxiv.org/abs/2106.02056}{\textcolor{black}{2106.02056}} using the GAMBIT framework. Here we show results only for the dimension-6 operators that describe the interactions between a gauge-singlet Dirac fermion and Standard Model quarks. Our global fit combines the latest constraints from \emph{Planck}, direct and indirect DM detection, and the LHC. For DM mass below 100 GeV, it is impossible to simultaneously satisfy all constraints while maintaining the EFT validity at high energies. For higher masses, however, large regions of parameter space remain viable where the EFT is valid and saturates the observed DM abundance.} 


\section{Introduction}\label{sec:intro}
The true particle nature of dark matter (DM) continues to remain a mystery. Many DM candidates have been proposed in the literature that fall in either a bottom-up, effective field theory (EFT) or top-down approach. In the former case, one includes a set of effective higher-dimensional operators to describe the interactions between DM and Standard Model (SM) particles at lower energy scales. Meanwhile, the latter case involves a high-energy, ultraviolet theory that allows one to make reliable predictions at all energy scales. 

Here we focus on the EFT approach and present a subset of our full results based on Ref.~\cite{GAMBIT:2021rlp} using the \textsf{GAMBIT} framework \cite{gambit}. These results are for a Dirac fermion DM particle $\chi$ that interacts with SM quarks via dimension-6 operators. In our global fit, we include contributions from the DM relic density, indirect and direct detection experiments, and the LHC. We also include likelihoods for 8 nuisance parameters to characterise the uncertainties associated with the top-quark running mass, nuclear form factors and astrophysical distribution of DM. This allows us to construct 2D profile likelihood plots in the relevant parameter planes of interest using a frequentist approach.

\section{Dirac fermion DM EFT}
We assume a Dirac fermion DM particle $\chi$ that interacts with SM quarks via dimension-6 effective, relativistic operators.\footnote{Here we focus only on dimension-6 operators. For our full set of results, including ones for dimension-6 and 7 operators, see Ref.~\cite{GAMBIT:2021rlp}.}~The model Lagrangian is given by \cite{GAMBIT:2021rlp}
\begin{equation}
	\lagr_\chi = \lagr_{\mathrm{SM}} + \overline{\chi} (i \slashed{\partial} - m_\chi) \chi + \lagr_{\mathrm{int}}\,,
\end{equation}
where the interaction Lagrangian $\lagr_{\mathrm{int}}$ is
\begin{align}
	\begin{split}
		\lagr_{\mathrm{int}} &= \frac{1}{\Lambda^2} \left[\mathcal{C}_1^{(6)}~(\overline{\chi} \gamma_\mu \chi) (\overline{q} \gamma^\mu q) +~\mathcal{C}_2^{(6)}~(\overline{\chi} \gamma_\mu \gamma_5 \chi) (\overline{q} \gamma^\mu q) \right. \\
		&\hspace{1.1cm} \left. +~\mathcal{C}_3^{(6)}~(\overline{\chi} \gamma_\mu \chi) (\overline{q} \gamma^\mu \gamma_5 q) +~\mathcal{C}_4^{(6)}~(\overline{\chi} \gamma_\mu \gamma_5 \chi) (\overline{q} \gamma^\mu \gamma_5 q) \right] \,.
	\end{split}
\end{align}
Here $\calC_i^{(6)}$ for $i=1,\ldots,4$ are dimensionless Wilson coefficients (for dimension-6 operators) defined at the new physics scale $\Lambda$. Thus, our model contains 6 free parameters ($\calC_{1 \ldots 4}^{(6)}$\,, $m_\chi$ and $\Lambda$).

\section{Constraints and likelihoods}
Under the renormalisation group flow, the Wilson coefficients run with energy and mix amongst each other. For instance, in computing direct detection limits, the Wilson coefficients are calculated at an energy scale $\mu = 2$\,GeV. In addition, when $\mu$ is above/below a quark mass (e.g., top quark), threshold corrections appear. Both the running and mixing of the Wilson coefficients as well as threshold corrections are taken into account using the \textsf{DirectDM\,v2.2.0} \cite{Bishara:2017nnn,Brod:2017bsw} package.

To impose constraints on the model parameter space, we construct a joint likelihood function. Here is a summary of various likelihood functions that enter in our fit.
\begin{enumerate}
	\item \textbf{Direct detection}: The Wilson coefficients are evaluated at $\mu = 2$\,GeV using \textsf{DirectDM} and matched onto a set of non-relativistic EFT operators. These are used in \textsf{DDCalc\,v2.2.0} \cite{DarkBit,HP} to compute predicted events rates and corresponding likelihoods for XENON1T, LUX (2016), PandaX (2016) and (2017), CDMSlite, CRESST-II and CRESST-III, PICO-60 (2017) and (2019), and DarkSide-50 experiments.
	
	\item \textbf{Relic density}: Using \textsf{CalcHEP\,v3.6.27} \cite{Belyaev:2012qa}, \textsf{GUM} \cite{GUM} and \textsf{DarkSUSY\,v6.2.2} \cite{darksusy}, we compute the DM relic density via a thermal freeze-out scenario. Both cases where $\chi$ makes up all ($f_\chi \equiv \Omega_\chi/0.12 \approx 1$) or a sub-component ($f_\chi \leq 1$) of the total DM abundance are studied.
	
	\item \textbf{\emph{Fermi}-LAT searches for gamma rays}: Observations of dwarf spheroidal galaxies of the Milky Way place strong constraints on the DM annihilation rate. Using \emph{Fermi}-LAT searches for gamma rays from DM annihilation in dwarfs \cite{LATdwarfP8}, we use the \textsf{gamLike\,v1.0.1} package within \textsf{DarkBit} \cite{DarkBit} to compute the resulting likelihood function.

	\item \textbf{Solar capture}: Neutrinos from DM annihilation in the Sun can be detected at the IceCube experiment. Using \textsf{Capt’n\,General} \cite{Kozar:2021iur}, we compute the DM capture rate in the Sun and utilise the \textsf{nulike} \cite{IC22Methods} package to obtain an event-by-event level likelihood for the 79-string IceCube data \cite{IC79_SUSY}. 
	
	\item \textbf{Energy injection bounds}: Using the \textsf{CosmoBit} \cite{GAMBITCosmologyWorkgroup:2020htv} module of \textsf{GAMBIT}, we compute bounds on our model based on predicted rates of DM annihilation in the early universe. These annihilations lead to energy injection \cite{Slatyer15a,Slatyer:2015kla} and observable effects in the cosmic microwave background.

	\item \textbf{ATLAS and CMS monojet searches}: By combining the \textsf{ColliderBit} \cite{ColliderBit} module of \textsf{GAMBIT} with \textsf{FeynRules\,v2.0} \cite{Alloul:2013bka}, \textsf{MadGraph\_aMC@NLO\,v2.6.6} \cite{Alwall:2011uj}, \textsf{Pythia\,v8.1} \cite{Sjostrand:2007gs} and \textsf{Delphes\,v3.4.2} \cite{DELPHES3}, we compute a likelihood based on monojet searches performed at the ATLAS and CMS experiments.
\end{enumerate}
In addition to 6 free model parameters, we also include 8 nuisance parameters (see above). This leads to a 14-dimensional parameter space.

We also ensure that the EFT remains valid for relevant energy scales of interest. This corresponds to
\begin{align}
	\Lambda \gtrsim 2\,\textrm{GeV}~~&(\textrm{direct~detection}), \\ 
	\Lambda \gtrsim 2 m_\chi~~&\textrm{(\textrm{relic density, indirect detection})}\,, \\
	\slashed{E}_T < \Lambda~~&\textrm{(\textrm{collider searches})}\,.
\end{align}
Here $\slashed{E}_T$ refers to the missing transverse energy. For collider searches, we modify the $\slashed{E}_T$ spectrum for $\slashed{E}_T > \Lambda$ via the following prescription:
\begin{equation}\label{eqn:sig_dist}
	\dfrac{d \sigma}{d \slashed{E}_T} \rightarrow
	\begin{cases}
		0\,, & \textrm{hard cut-off}\,, \\[1mm]
		\dfrac{d \sigma}{d \slashed{E}_T} \left(\dfrac{\slashed{E}_T}{\Lambda} \right)^{-a}\,, & \textrm{smooth cut-off}\,.
	\end{cases}
\end{equation}
Here $a \in [0, 4]$ is a nuisance parameter that is profiled over in our study.

By separating the scale of new physics from the Wilson coefficient, we can impose the perturbativity bound: $|\calC_i^{(6)}| < 4\pi$. As for the scan range of model parameters, we use the following values:
\begin{equation}
	m_\chi \in [5,\,500]\,\textrm{GeV}\,, \quad \Lambda \in [20,\,2000]\,\textrm{GeV}\,.
\end{equation}
We adopt a frequentist approach and profile over the nuisance parameters in our study. This allows us to construct 2D profile likelihood plots for relevant model parameters at 68.3\% (1$\sigma$) and 95.4\% (2$\sigma$) confidence level.

\section{Results}
\begin{figure}[t]
	\centering
	
	\includegraphics[width=0.48\textwidth]{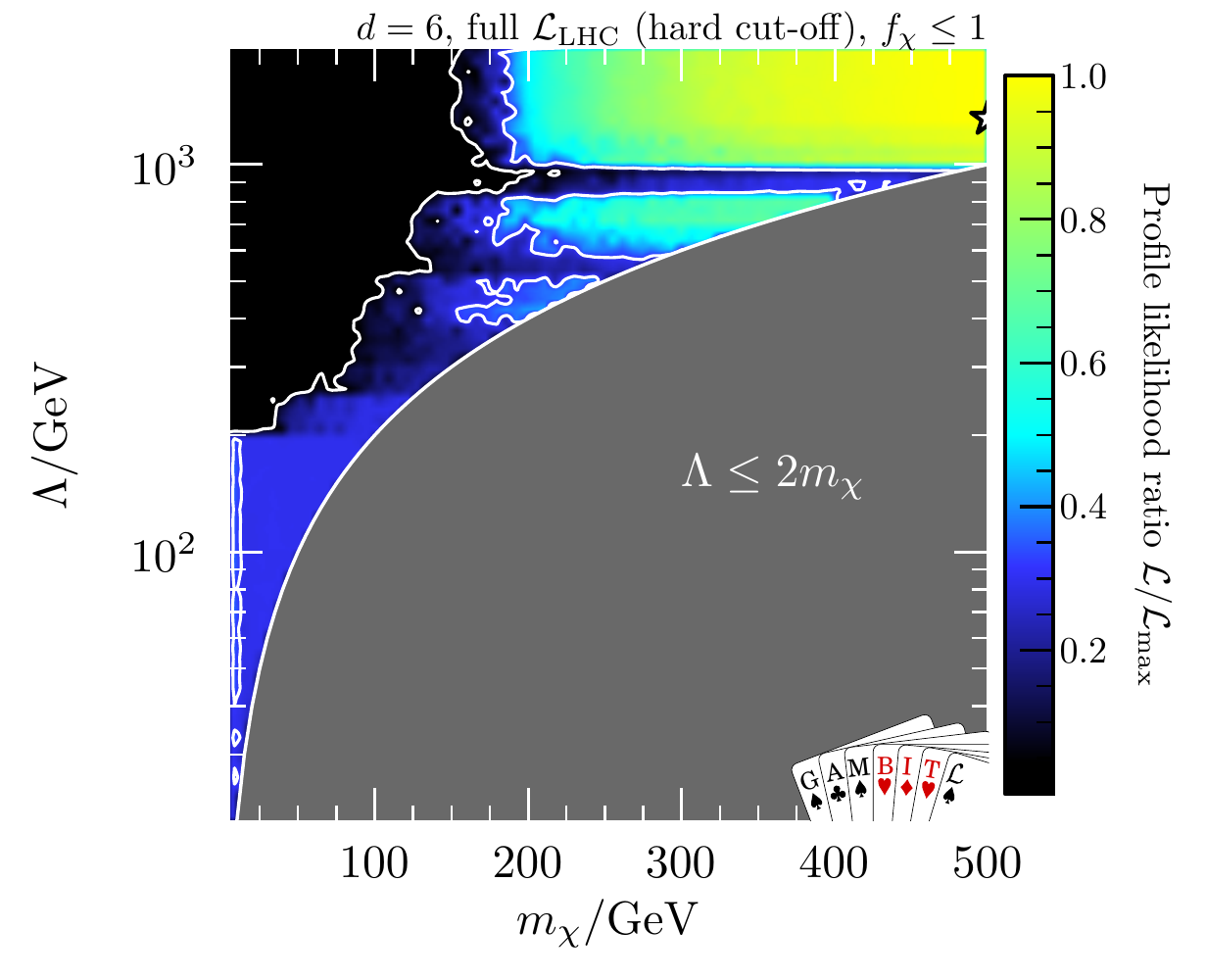} \quad
	\includegraphics[width=0.48\textwidth]{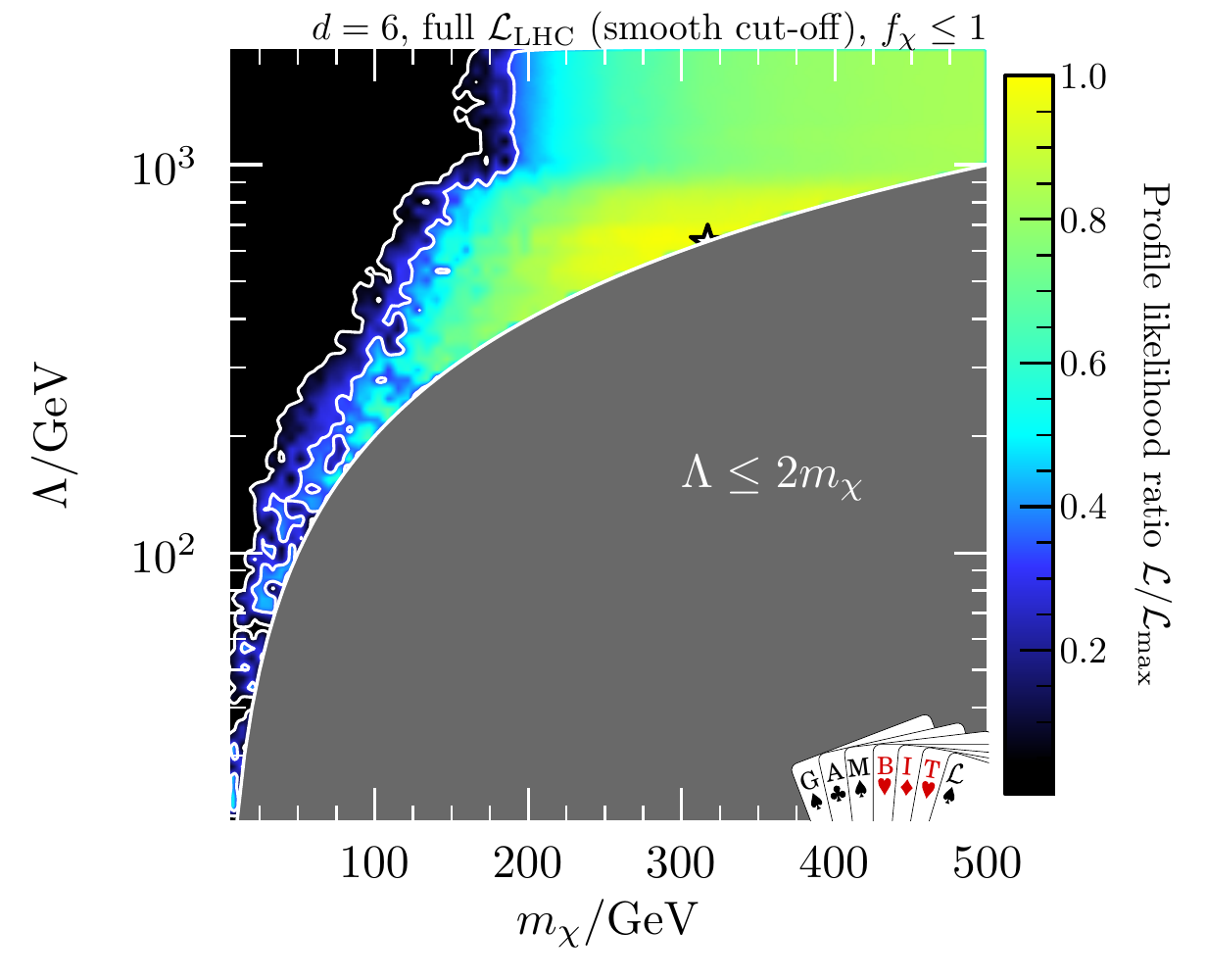}
	
	\caption{2D profile likelihood plots in the $(m_\chi,\Lambda)$ plane for the full LHC likelihood with hard (\emph{left panel}) and smooth (\emph{right panel}) cut-off in the missing transverse energy spectrum. The best-fit point is shown by the white star, whereas the solid white lines show the 68.3\% (1$\sigma$) and 95.4\% (2$\sigma$) confidence level contours. The grey region is ruled out by the EFT validity requirement $(\Lambda > 2m_\chi)$ for the relic density and indirect detection limits.}
	\label{fig:mX_lambda_plane}
\end{figure}

In Fig.~\ref{fig:mX_lambda_plane}, we show the 2D profile likelihood plots in the $(m_\chi,\,\Lambda)$ plane using the full LHC likelihood with hard (\emph{left panel}) and smooth (\emph{right panel}) cut-off in the $\slashed{E}_T$ spectrum. The best-fit point is depicted by a white star, whereas the solid white lines correspond to 1 and 2$\sigma$ confidence level contours. 

The grey region is ruled out by the EFT validity requirement for the DM relic density and indirect detection limits. For $m_\chi \lesssim 100$\,GeV, it is impossible to obtain the correct abundance of $\Omega_\chi h^2 = 0.12$ due to combined constraints from indirect and direct detection experiments. For $m_\chi > 100$\,GeV and $\Lambda > 200$\,GeV, the LHC constraints are strong enough in the sub-component DM case $(f_\chi \leq 1)$ that the requirement from relic density rules out most of the parameter space. These constraints are weaker for $m_\chi \gtrsim 1$\,TeV as LHC energy becomes insufficient to produce heavy DM particles.

In the left panel, we see two preferred regions in the parameter space for $\Lambda \approx 700$\,GeV (CMS) and $\Lambda \gtrsim 1$\,TeV (ATLAS). These two regions appear as a result of small excesses seen in the few high-$\slashed{E}_T$ bins of the CMS and ATLAS monojet searches respectively.\footnote{We also find similar results in the case of combined dimension-6 and 7 operators, see Fig.~11 in Ref.~\cite{GAMBIT:2021rlp}.} Meanwhile, the effect of using a smooth cut-off in the $\slashed{E}_T$ spectrum for $\slashed{E}_T > \Lambda$ is evident in the right panel. Here the best-fit solution gives a better fit to both excesses simultaneously and prefers $a \approx 1.7$ in Eq.~\eqref{eqn:sig_dist}.

\section{Summary}
We have performed a first global analysis of full set of effective operators for a Dirac fermion DM interaction with SM quarks. Using a novel approach to address the issue of EFT validity at the LHC, we used highly efficient likelihood calculations and sampling algorithms to sample the 14-dimensional model parameter space. 

Our results lead to strong constraints on small $m_\chi$ and large $\Lambda$. A slight preference for a DM signal was found at relatively small $\Lambda$. For the model to be compatible with the LHC constraints, we required $\Lambda \lesssim 200$\,GeV for $m_\chi \lesssim 100$\,GeV. However, large regions of the parameter space still remains viable whenever $\chi$ makes up all of the observed DM abundance.

\vspace{2mm}
\noindent \textbf{Note}: \emph{All of our results, samples and input files are publicly available via} \href{https://www.zenodo.org/record/4836397#.Yt6_qC8w2aM}{\textsf{Zenodo}}.

\section*{Acknowledgements}
A.B. gratefully acknowledges all members of the GAMBIT collaboration for their involvement, and the IDM 2022 conference organisers/convenors for the opportunity to present this work. 

\paragraph{Funding information} A.B. is supported by the Science and Technology Facilities Council (STFC) Grant No.~ST/T00679X/1.

\bibliography{IDM_2022_AB.bib}
\bibliographystyle{JHEP}

\nolinenumbers

\end{document}